\documentclass{article}

\usepackage{graphicx}
\usepackage{amsmath, amssymb}
\usepackage{amsthm}
\usepackage{bm}

\theoremstyle{definition}

\theoremstyle{plain}
\newtheorem{theorem}{Theorem}[section]

\title{Diffusive Limit of Hawkes Driven Order Book Dynamics With Liquidity Migration}
\author{Levon Sarkis Mahseredjian}
\date{October 2025}

\begin{document}

\maketitle

\begin{abstract}
This paper develops a theoretical mesoscopic model of the limit order book driven by multivariate Hawkes processes, designed to capture temporal self-excitation and the spatial propagation of order flow across price levels. In contrast to classical \textit{zero-intelligence} or Poisson based queueing models, the proposed framework introduces mathematically defined migration events between neighbouring price levels, whose intensities are themselves governed by the underlying Hawkes structure. This provides a principled stochastic mechanism for modeling interactions between order arrivals, cancellations, and liquidity movement across adjacent queues.

Starting from a microscopic specification of Hawkes driven order flow, we derive a diffusion approximation which yields a reflected mesoscopic stochastic differential equation (SDE) system for queue volumes. The limiting generator is obtained through a Taylor expansion of the microscopic generator, demonstrating how temporal excitation together with spatial migration determine the drift and diffusion structure of the limit order book in the mesoscopic regime. The resulting model extends existing diffusion limits by incorporating correlated excitations and price level to price level liquidity movement within a unified Hawkes based formulation.

By establishing this diffusive limit, the paper provides a mathematically consistent bridge between high frequency event based models and macroscopic stochastic descriptions of market microstructure. The work is entirely theoretical and lays a foundation for future analytical and numerical developments without relying on empirical calibration.
\end{abstract}

\noindent\textbf{Keywords:} Limit Order Book, Hawkes Processes, Functional Central Limit Theorem, Market Microstructure, Stochastic Modelling.
\newpage
\tableofcontents

\clearpage

\section{Introduction}

The limit order book (LOB) is a discrete queuing system that records all outstanding buy and sell orders awaiting execution. Each order specifies both a price and a quantity, and transactions occur when compatible buy and sell orders meet. The limit order book thus provides a detailed and dynamic representation of market liquidity.
\paragraph{}
In this work, we focus on the \textit{first-in, first-out} (FIFO) queuing rule, which prioritises orders by their time of submission. Although alternative, more complicated, mechanisms such as Pro-Rata exist, the FIFO structure remains the standard in most modern electronic exchanges. Modelling the LOB is a central problem in computational finance, as it enables the study of market microstructure, price formation, and execution dynamics. A significant class of limit order book models emerges from the \textit{perfect-rationality} approach, in which market participants are assumed to behave strategically and optimise their order placement decisions. These models treat order submission, cancellation, and execution as the outcome of agents maximising expected utility under market frictions and informational constraints. This framework, pioneered in the economics literature, provides a microstructure consistent explanation for observed order book shapes and trading behaviours. A comprehensive survey of these models is provided by Parlour and Seppi [2]. Institutions such as hedge funds and market making firms regularly use such models to design and test algorithmic trading strategies. In contrast, the \textit{zero-intelligence} approach models the limit order book as a purely stochastic system, where order arrivals, cancellations, and executions are treated as exogenous random events rather than the outcome of strategic optimisation. This framework replaces behavioural assumptions with mathematically tractable probabilistic structures, often using Poisson processes or Markovian queueing dynamics to describe the evolution of liquidity across price levels. We adopt a \textit{zero-intelligence} modelling framework, where order flow is represented as a sequence of random events rather than as the outcome of perfectly rational agent decisions. This approach is utilised in this paper and allows for analytical tractability while still reproducing key empirical regularities observed in high frequency market data.

\subsection{Limit Order Book}
A limit order book operates as a double auction mechanism in which buy and sell orders are matched based on price and time via a matching algorithm. The \textit{bid side} aggregates all buy limit orders, arranged in descending order of price, while the \textit{ask side} aggregates all sell limit orders, arranged in ascending order of price. The highest bid and the lowest ask define the \textit{best bid} and \textit{best ask} prices, respectively, and their difference is known as the \textit{bid ask spread} which is what market makers and other market participants seek to profit from.

New orders arriving to the book may either: add liquidity (limit orders) by joining the queue at a given price level;
remove liquidity (market orders) by executing immediately against standing orders; or
cancel existing liquidity(cancellations), thereby modifying the depth at a price level.

The stochastic evolution of these order types governs both price movements and liquidity fluctuations. Consequently, realistic LOB models must capture the statistical dependencies between such events. This motivates the use of Hawkes processes, as a more effective point process for modeling the self-exciting and mutually-exciting nature of high frequency order flow which has been shown empirically is many such papers, for instance in [3]. Another important feature of the limit order book is the \textit{tick size}. A \emph{tick} is the minimum allowable price increment in the order book. The bid ask spread is always an
integer number of ticks.

\subsection{Literature Review}

The modeling of limit order book dynamics has undergone a substantial evolution over the past two decades. Early stochastic models such as those developed by Cont and de Larrard [4] established that simple queueing mechanisms with Poisson order flow can reproduce several stylised empirical features of electronic markets, including average depth profiles and the distribution of times to price changes. These models offered analytical tractability and clear structural interpretation, but the assumption of independent Poisson arrivals limited their capacity to represent the pronounced temporal clustering observed in high frequency markets.
\paragraph{}
A major advancement came from the introduction of multivariate Hawkes processes. The empirical work of Bacry, Muzy and collaborators [5] demonstrated that order submissions, cancellations and market orders exhibit strong self-excitation and cross-excitation, revealing that modern markets are highly endogenous systems in which activity propagates across event types and between buy and sell sides. This insight motivated a new generation of microscopic order book models in which each event type is governed by a Hawkes intensity.
The question of how such microscopic Hawkes driven systems behave under scaling limits has been addressed in several recent studies. The work of Horst and Paul [6] considered a high frequency limit in which queue sizes are rescaled diffusively. They showed that a Hawkes driven birth death mechanism at each price level converges to a reflected diffusion whose drift and volatility depend explicitly on the underlying Hawkes kernels. Their framework provides a rigorous bridge between self-exciting microscopic order flow and mesoscopic stochastic differential equations. Hambly and Ledger [7] pursued a complementary perspective by studying the behaviour of the order book when the number of price levels grows and the tick size tends to zero. Their analysis led to reflected stochastic partial differential equations describing the evolution of the order book as a continuum in price space. Together, these works reveal the usefulness of scaling limits in connecting high frequency market microstructure with more tractable diffusion style models.
\paragraph{}
Additional insight into Hawkes modulated order flow is provided by the doctoral work of Chen [8], who investigated the interaction between excitation in order flow and short term price predictability. Chen’s analysis highlighted the role of cross-excitation between order types and the relationship between local imbalance and subsequent price movements. This empirical and modelling evidence further strengthens the argument that excitation must be incorporated directly into realistic models of queue dynamics.

Earlier theoretical work, such as that of Abergel et al [9], studied the propagation of liquidity and the relationship between order flow events and price formation from a more structural perspective. Although these models were not constructed around Hawkes processes, they emphasised the need for a mechanistic understanding of how order book events at one level influence neighbouring levels and eventually drive observable price changes.

The present work builds on and extends these strands of literature. In contrast to existing mesoscopic Hawkes driven models, which primarily treat each queue level as an isolated birth death process, I incorporate a microscopic mechanism for liquidity migration between neighbouring price levels. This migration is governed by Hawkes modulated intensities and, under diffusive scaling, produces a discrete Laplacian term in the limiting generator. The resulting mesoscopic dynamics therefore combine temporal self-excitation with a diffusion like spatial coupling between queue levels, reflecting the empirically observed spread of liquidity across adjacent depths. Furthermore, by deriving the limiting generator explicitly and connecting the diffusion covariance matrix to the Hawkes branching structure through a functional central limit theorem, my model retains a direct and interpretable link between microscopic excitation and macroscopic variability. Existing models either omit spatial liquidity propagation or introduce it in an ad hoc manner, whereas in my framework it arises naturally from the underlying Hawkes structure.

By integrating endogenous Hawkes excitation with liquidity migration and by establishing a rigorous diffusive limit that yields a reflected multi-dimensional SDE with Laplacian coupling, this work advances the literature toward a more complete and empirically faithful description of limit order book dynamics. It offers a mesoscopic model that remains grounded in microscopic market mechanics while providing a tractable analytical structure suitable for theoretical analysis, simulation and empirical calibration.

\section{Hawkes Driven Order Flow}

In this section, we introduce the Hawkes process, a stochastic framework designed to capture the self-exciting and mutually-exciting dynamics of event arrivals in high frequency financial markets.

Through this formulation, Hawkes processes provide a flexible framework to represent key stylised facts of limit order book dynamics. They naturally capture self-excitation, whereby a surge in buy (or sell) orders raises the short-term probability of additional activity of the same type; cross-excitation, where actions on one side of the book influence behaviour on the opposite side (for instance, a sell side cancellation may encourage more buy side limit orders); and temporal decay, reflecting the gradual fading of such effects over a characteristic timescale. 

These mechanisms collectively explain a range of empirical phenomena observed in high frequency data, including volatility clustering, order flow imbalance persistence, and long range dependence in event arrivals. Consequently, Hawkes processes serve as a natural and tractable foundation for modeling order flow in both microscopic and mesoscopic representations of the limit order book.

\subsection{Point Processes and Conditional Intensity}
The definitions in this section follow standard Hawkes process theory, I refer readers to [4] [5] if they seek more detail.

A point process on $\mathbb{R}_+$ is a sequence of random event times $(T_k)_{k\in\mathbb{N}}$ such that $0 < T_1 < T_2 < \dots$. The associated \textit{counting process} is defined by
\begin{equation}
N(t) = \sum_{k \geq 1} \mathbf{1}_{\{T_k \leq t\}}, \quad t \geq 0,
\end{equation}
which counts the number of events up to time $t$.  
The \textit{conditional intensity function} $\lambda(t)$ characterizes the instantaneous event rate given the past filtration $\mathcal{F}_t$:
\begin{equation}
\lambda(t) := \lim_{\Delta \to 0} \frac{\mathbb{E}[N(t+\Delta) - N(t) \mid \mathcal{F}_t]}{\Delta}.
\end{equation}
This quantity determines the expected rate of arrivals at time $t$ conditional on all prior events.
A univariate Hawkes process is a point process $N(t)$ whose intensity depends linearly on its own history:
\begin{equation}
\lambda(t) = \mu + \int_0^t \phi(t-s)\, dN(s),
\label{eq:hawkes_uni}
\end{equation}
where $\mu > 0$ is the \textit{baseline intensity}, and $\phi : \mathbb{R}_+ \to \mathbb{R}_+$ is the \textit{kernel function} describing how past events excite future ones.  
Intuitively, each arrival increases the conditional intensity for a period of time determined by the kernel $\phi$. A common choice is the exponential kernel,
\begin{equation}
\phi(t) = \alpha e^{-\beta t}, \quad \alpha, \beta > 0,
\label{eq:exp_kernel}
\end{equation}
which implies that excitation decays exponentially with rate $\beta$. The parameters $\alpha$ and $\beta$ respectively control the magnitude and persistence of self-excitation.

\subsection{Multivariate Hawkes Process}

To capture interactions between different types of events (e.g., buy/sell, limit/cancel/market orders), the model is extended to a $d$-dimensional vector of counting processes $N(t) = (N_1(t), \dots, N_d(t))$. Each component has its own intensity:
\begin{equation}
\lambda_i(t) = \mu_i + \sum_{j=1}^d \int_0^t \phi_{ij}(t-s)\, dN_j(s), 
\quad i = 1, \dots, d,
\label{eq:multi_hawkes}
\end{equation}
where $\mu_i$ is the baseline intensity of process $i$, and $\phi_{ij}$ encodes how events of type $j$ influence the future rate of type $i$ events.  
The matrix $\Phi(t) = [\phi_{ij}(t)]_{i,j=1}^d$ is called the \textit{kernel matrix}, and the process can be compactly written as
\begin{equation}
\lambda(t) = \mu + \int_0^t \Phi(t-s)\, dN(s).
\end{equation}

\subsection{Stationarity and Stability}

Define the integrated kernel matrix $K = \int_0^\infty \Phi(s)\, ds$.  
The multivariate Hawkes process is stable and stationary if the spectral radius $\rho(K) < 1$.  
In that case, the expected intensity vector is
\begin{equation}
\Lambda = (I - K)^{-1}\mu,
\end{equation}
where $\Lambda_i = \mathbb{E}[\lambda_i(t)]$ is the long run mean event rate of type $i$.  
This condition ensures that excitation remains bounded and that event clusters do not lead to explosive growth in intensity.

\section{Microscopic Model of The Limit Order Book  (Static Setting)}
\label{sec:micro-static}

We model each side of the limit order book as a vector of integer valued queues indexed
by their distance (in ticks) from the mid. Throughout this section, the mid price is held fixed
and the bid ask spread is assumed to be two ticks. Let $N\in\mathbb{N}$ be the maximum depth
we track on each side (in ticks from the mid, excluding the best quote queues themselves).

\subsection{State, indexing and unit vectors}
\label{subsec:state-indexing}
While our definition of the microscopic structure remains standard our contribution appears in the mesoscopic scaling limit, a Laplacian smoothing term across price levels. We begin by fixing the bid ask spread to be constantly equal to two ticks. This assumption is
appropriate for large tick assets, where the best bid and best ask typically remain stably one
tick away from the mid price. Mathematically, a fixed spread ensures that the queue indices
correspond to fixed tick distances from the mid price, so that the microscopic state space has
constant dimension. Without this assumption, the best bid and best ask queues would shift
position as the spread fluctuates, causing the queue vectors to change dimension over time and
violating the regularity and tightness conditions required for the scaling limit theorem.
Maintaining a constant spread therefore guarantees that the rescaled birth death dynamics evolve
on a stable, fixed coordinate system, which is essential for deriving a diffusion-type limit. We consider two classes of queue vectors, one for the bid side and one for the ask side.
For the bid side, define
\[
    Z^b_n(t)
    = \big( Z^{b,1}_n(t),\ldots,Z^{b,N-1}_n(t) \big)
    \in \mathbb{Z}_+^{\,N-1}.
\]

For the ask side, define
\[
    Z^a_n(t)
    = \big( Z^{a,1}_n(t),\ldots,Z^{a,N-1}_n(t) \big)
    \in \mathbb{Z}_+^{\,N-1}.
\]
Here $Z^{b,i}_n(t)$ (resp.\ $Z^{a,i}_n(t)$) denotes the number of outstanding buy (resp.\ sell)
limit orders posted at the price level $m - i$ (resp.\ $m + i$), where $m$ is the mid price. We
assume that all order and cancellation sizes equal one. At any price level $i$ on the bid side,
the quantity $Z^{b,i}_n(t)$ increases by one upon the arrival of a buy limit order and decreases
by one following the arrival of a sell market order or a cancellation of an existing buy limit
order. Analogous dynamics hold for $Z^{a,i}_n(t)$ on the ask side.

For $i \in \{1,\ldots,N-1\}$, we let $e_i \in \mathbb{R}^{N-1}$ denote the $i$-th standard basis
vector.

\subsection{Microscopic Dynamics}

We now describe the microscopic dynamics governing the bid-side queue
process $Z^b_n = (Z^{b,1}_n,\dots,Z^{b,N-1}_n)$.  
At the discrete level, the volume at each price level evolves through a 
collection of elementary jumps corresponding to order arrivals, cancellations,
and reallocations of volume between neighbouring price levels.  

Each of these elementary mechanisms is represented by an \emph{independent}
Poisson process whose intensity may depend on the current state of the order
book.  We denote by $e_i$ the $i$-th canonical basis vector of
$\mathbb{R}^{N-1}$, and we adopt the convention $e_0 = e_N = 0$.
\paragraph{Bid-side dynamics.}
For the bid side, the microscopic transitions are as follows:
\paragraph{1. Limit order arrivals.}
For every level $i \in \{1,\dots,N-1\}$, the queue increases by one unit,
\[
 Z^b_n \longrightarrow Z^b_n + e_i,
\]
at the arrival rate according 
according to the Hawkes process $N^1_{b,n,i}(t)$ with intensity
\begin{align}
\lambda^{1}_{b,n,i}(t)
&= \mu^{1}_{b,n,i}\!\big(Z^{b,i}_n(t^-)\big)
  + \int_0^t \alpha^{11}_{b,n,i} e^{-\beta^{11}_{b,n,i}(t-s)}\, dN^{1}_{b,n,i}(s)
  + \int_0^t \alpha^{12}_{b,n,i} e^{-\beta^{12}_{b,n,i}(t-s)}\, dN^{2}_{b,n,i}(s)
\nonumber\\[-0.25em]
&\hspace{7.25em}
  + \int_0^t \alpha^{14}_{b,n,i} e^{-\beta^{14}_{b,n,i}(t-s)}\, dN^{4}_{b,n,i}(s).
\label{eq:hawkes-bid-birth}
\end{align}
\paragraph{2. Cancellations and market order removals.}
For every level $i \in \{1,\dots,N-1\}$, one unit is removed from the queue,
\[
 Z^b_n \longrightarrow Z^b_n - e_i,
\]
at the arrival rate according to the Hawkes process $N^2_{b,n,i}(t)$ with intensity
\begin{align}
\lambda^{2}_{b,n,i}(t)
&= \Bigg(
     \mu^{2}_{b,n,i}\!\big(Z^{b,i}_n(t^-)\big)
     + \int_0^t \alpha^{21}_{b,n,i} e^{-\beta^{21}_{b,n,i}(t-s)}\, dN^{1}_{b,n,i}(s)
     + \int_0^t \alpha^{22}_{b,n,i} e^{-\beta^{22}_{b,n,i}(t-s)}\, dN^{2}_{b,n,i}(s)
   \Bigg)\,\mathbf{1}_{\{Z^{b,i}_n(t^-)>0\}}.
\label{eq:hawkes-bid-death}
\end{align}

\paragraph{3. Volume reallocation: migration from level $i$ to level $i-1$.}
Each unit of volume sitting at level $i$ independently attempts to jump one
tick closer to the midprice.  Consequently, the queue updates according to
\[
Z^b_n \longrightarrow Z^b_n + e_{i-1} - e_i,
\]
the total migration rate is Hawkes modulated and proportional to the
current queue size:
\[
    \lambda^{i\to i-1}_{b,n}(t)
    \;=\; a^{b,i}_n(t)\, Z^{b,i}_n(t-),
\]
where the symmetric Hawkes per unit migration intensity is given by
\[
    a^{b,i}_n(t)
    \;=\;
    \eta^{b,i}_n
    + \sum_{\ell \in \{1,2,4\}}
        \int_0^t 
        \kappa^{b,i}_\ell \, e^{-\rho^{b,i}_\ell (t-s)} \, dN^{\ell}_{b,n,i}(s).
\]
Here $\eta^{b,i}_n \ge 0$ is the baseline per unit migration rate and
$(\kappa^{b,i}_\ell, \rho^{b,i}_\ell)$ are the Hawkes excitation parameters.
Because migration is symmetric, the same $a^{b,i}_n(t)$ governs both inward
and outward jumps.

\paragraph{4. Volume reallocation: migration from level $i$ to level $i+1$.}
Similarly, each unit of volume may jump one tick further from the midprice.
The corresponding transition is
\[
Z^b_n \longrightarrow Z^b_n + e_{i+1} - e_i,
\]
occurring at total rate
\[
    \lambda^{i\to i+1}_{b,n}(t)
    \;=\; a^{b,i}_n(t)\, Z^{b,i}_n(t-),
\]
with the same symmetric Hawkes intensity $a^{b,i}_n(t)$ as above.  
Symmetry means that the per-unit migration rate towards and away from 
the midprice is identical:
\[
    a^{b,i,+}_n(t) = a^{b,i,-}_n(t) = a^{b,i}_n(t).
\]
Thus, migration contributes a discrete Laplacian smoothing effect at the
mesoscopic scale while allowing the strength of reallocation to be 
 history dependent through Hawkes excitation

These transition rules fully characterise the microscopic bid queue dynamics.
The ask side of the book is defined analogously.

\section{Mesoscopic Model Of the Limit Order Book}

\subsection{Scaling Limit of the Microscopic Model}

We now derive the mesoscopic limit of the microscopic Hawkes driven order book. 
Following the scaling arguments of Hambley et al [1], we accelerate time by a
factor of $n$ and renormalise queue sizes by $\sqrt{n}$.
For the bid side, define the rescaled processes
\[
    \widetilde{Z}^{\,b}_n(t)
    := \frac{Z^{\,b}_n(nt)}{\sqrt{n}}
    \qquad\text{and}\qquad
    \widetilde{Z}^{\,a}_n(t)
    := \frac{Z^{\,a}_n(nt)}{\sqrt{n}}.
\]
The limiting processes will form the mesoscopic (continuous volume,
discrete‐price) limit order book.

\subsection{Scaling assumptions}

We impose the following standard conditions, adapted to the Hawkes setting:

\begin{itemize}
    \item[(i)] \textbf{Hawkes stability.}  
    The spectral radius of the Hawkes kernel matrix is strictly less than~1,
    uniformly in~$n$, ensuring non-explosion and existence of stationary 
    intensities.

    \item[(ii)] \textbf{Regularity of baselines.}  
    The baseline functions for arrivals and cancellations satisfy global 
    Lipschitz and linear growth bounds in the queue state.

    \item[(iii)] \textbf{Moment bounds.}  
    The initial states satisfy 
    $\sup_{n} \mathbb{E}\|X^{k}_n(0)\|^2 < \infty$ 
    for $k\in\{b,a\}$.

    \item[(iv)] \textbf{Predictability and conditional independence.}  
    All microscopic intensities depend on the predictable left-limit 
    $Z^{k}_n(t-)$, and conditioned on the filtration $\mathcal{F}_{t-}$—the 
    drivers (arrival, cancellation/market, and migration processes) are 
    independent.

    \item[(v)] \textbf{Boundary behaviour.}  
    Migration at the boundary levels is blocked consistently with the pinning 
    conditions, and Skorokhod reflection is used to preserve non-negativity.
\end{itemize}

These conditions parallel those in Hambly et al [1].

\subsection{Limit dynamics}

Let
\[
    A^{k,i}_t := \lim_{n\to\infty} a^{k,i}_n(t)
\]
denote the limiting Hawkes migration intensities. 
Let $h_{k,m}(i,x)$ denote the drift contribution arising from Hawkes driven 
arrivals and cancellations/market orders, and let $\sigma^{k}$ be the diffusion 
coefficient obtained from the limiting predictable quadratic variations.

The next result gives the mesoscopic reflected diffusion approximation.

\begin{theorem}[Mesoscopic Reflected SDE Limit]
\label{thm:mesoscopic}
Under the assumptions above, the sequence 
$(X^{b}_n, X^{a}_n)$ is tight in 
$\mathbb{D}([0,\infty);\mathbb{R}^{N-1})^2$, 
and any weak limit $(X^{b},X^{a})$ is the unique 
strong Markov process satisfying, for $i=1,\dots,N-1$,
\begin{equation}\label{eq:mesoscopic-bid}
\boxed{
\begin{aligned}
dX^{b,i}_t
&=
\Big[
    A^{b,i}_t\,
    \big(
        X^{b,i+1}_t
      + X^{b,i-1}_t
      - 2 X^{b,i}_t
    \big)
    + h_{b,m}\!\big(i,X^{b,i}_t\big)
\Big] dt
+ \sum_{j=1}^{N-1}\sigma^{b}_{ij}\!\big(X_t,A_t\big)\, dW^{b,j}_t
+ d\eta^{b,i}_t, 
\end{aligned}}
\end{equation}
and
\begin{equation}\label{eq:mesoscopic-ask}
\boxed{
\begin{aligned}
dX^{a,i}_t
&=
\Big[
    A^{a,i}_t\,
    \big(
        X^{a,i+1}_t
      + X^{a,i-1}_t
      - 2 X^{a,i}_t
    \big)
    + h_{a,m}\!\big(i,X^{a,i}_t\big)
\Big] dt
+ \sum_{j=1}^{N-1}\sigma^{a}_{ij}\!\big(X_t,A_t\big)\, dW^{a,j}_t
+ d\eta^{a,i}_t,
\end{aligned}}
\end{equation}
where $W^{b}$ and $W^{a}$ are Brownian motions with covariance determined by the
limiting intensities, and $\eta^{k,i}_t$ are reflection terms enforcing
$X^{k,i}_t\ge 0$ and $X^{k,0}_t=X^{k,N}_t=0$.
\end{theorem}

Theorem~\ref{thm:mesoscopic} shows that symmetric Hawkes migration contributes a 
discrete Laplacian smoothing term whose strength is 
history dependent through the Hawkes excitation dynamics.  
The remaining drift and diffusion coefficients arise from the Hawke
arrival, cancellation, and market order flows.  
Full proofs are provided in section 5.

\section{Mesoscopic model}

These events are driven by an underlying multivariate Hawkes process
$N_n(t)$ whose components correspond to the different event types. Denote by
$\lambda^{\mathrm{up}}_{n,i}(t)$, $\lambda^{\mathrm{down}}_{n,i}(t)$,
$\lambda^{\mathrm{mig},+}_{n,i}(t)$ and $\lambda^{\mathrm{mig},-}_{n,i}(t)$
the corresponding conditional intensities. We assume that, under an appropriate
equilibrium scaling regime, they admit the expansions
\begin{align}
  \lambda^{\mathrm{up}}_{n,i}(t)
  &= \frac{n}{2}\,\sigma_i^2\big(\tilde Z^b_{n,i}(t)\big)
     + \sqrt{n}\,f_i\big(\tilde Z^b_{n,i}(t)\big) + o(\sqrt{n}), \label{eq:up-intensity}\\
  \lambda^{\mathrm{down}}_{n,i}(t)
  &= \frac{n}{2}\,\sigma_i^2\big(\tilde Z^b_{n,i}(t)\big)
     + \sqrt{n}\,g_i\big(\tilde Z^b_{n,i}(t)\big) + o(\sqrt{n}), \label{eq:down-intensity}\\
  \lambda^{\mathrm{mig},\pm}_{n,i}(t)
  &= \sqrt{n}\,\alpha_b\,\tilde Z^b_{n,i}(t) + o(\sqrt{n}), \label{eq:mig-intensity}
\end{align}
where the effective coefficients $\sigma_i^2$, $f_i$, $g_i$ and $\alpha_b$ are
determined by the Hawkes kernels (see Section~\ref{sec:hawkes-covariance}).

\subsection{Diffusive rescaling}

We define
the rescaled process as in the previous section
\[
  \tilde Z^b_n(t) := \frac{Z^b_n(nt)}{\sqrt{n}},
  \qquad t \ge 0.
\]
The state space of $\tilde Z^b_n$ is the lattice
$\frac{1}{\sqrt{n}}\mathbb{N}^{N-1} \subset \mathbb{R}_+^{N-1}$, and each jump
of the microscopic process corresponds to a displacement of size $1/\sqrt{n}$
in one or two coordinates of $\tilde Z^b_n$.

\subsection{Generator of the rescaled process}

For a bounded function $F :
\frac{1}{\sqrt{n}}\mathbb{N}^{N-1} \to \mathbb{R}$, the infinitesimal
generator $A_n$ of the rescaled process $\tilde Z^b_n$ is given by
\begin{equation}
A_n F(y)
= \sum_{\Delta}
   \text{rate}(y \to y+\Delta)\,
   \big(F(y+\Delta) - F(y)\big)
\label{eq:An-generator}
\end{equation}

where the sum runs over all admissible jump vectors
$\Delta \in \frac{1}{\sqrt{n}}\{0,\pm e_i,\pm e_i \pm e_j\}$.

To write $A_n$ in a form amenable to Taylor expansion, we introduce the
finite difference operators
\begin{align*}
  \Delta^r_{n,k}F(y)
  &:= \sqrt{n}\,\big(F(y + e_k/\sqrt{n}) - F(y)\big),\\
  \Delta^l_{n,k}F(y)
  &:= \sqrt{n}\,\big(F(y) - F(y - e_k/\sqrt{n})\big),\\
  \Delta^2_{n,k}F(y)
  &:= n\,\big(F(y + e_k/\sqrt{n}) + F(y - e_k/\sqrt{n}) - 2F(y)\big).
\end{align*}
Using the microscopic dynamics and the intensity expansions (10), one checks that
$A_n F(y)$ can be written as
\begin{align}
  A_n F(y)
  &= \sum_{k=1}^{N-1}
     \frac{1}{2}\,\Delta^2_{n,k}F(y)\,\sigma_k^2(y_k)\,
       \mathbf{1}_{\{y_k \ge 1/\sqrt{n}\}} \nonumber\\
  &\quad + \sum_{k=1}^{N-1}
     \Delta^r_{n,k}F(y)\,\frac{\sigma_k^2(y_k)}{\sqrt{n}}\,
       \mathbf{1}_{\{y_k = 0\}} \nonumber\\
  &\quad + \sum_{k=1}^{N-1}
     \Big[
       \Delta^r_{n,k}F(y)\,f_k(y_k)
       - \Delta^l_{n,k}F(y)\,g_k(y_k)\,
         \mathbf{1}_{\{y_k \ge 1/\sqrt{n}\}}
     \Big] \nonumber\\
  &\quad + \sum_{k=1}^{N-1}
     \alpha_b\,y_k\,
     \Big(
       \Delta^r_{n,k-1}F(y - e_k/\sqrt{n})
       + \Delta^r_{n,k+1}F(y - e_k/\sqrt{n})
       - 2\Delta^l_{n,k}F(y)
     \Big).
  \label{eq:An-generator}
\end{align}

with the conventions $y_0 = y_N = 0$ and $\Delta^r_{n,0}F = \Delta^r_{n,N}F
\equiv 0$.

\paragraph{Remark 5.2.1}
The representation of the microscopic generator in (11)
and the rescaling procedure follow the classical diffusion approximation
framework for density dependent Markov jump processes developed by Ethier
and Kurtz; see [10].

\subsection{Limit generator via Taylor expansion}

Let $F \in C^2_b(\mathbb{R}_+^{N-1})$ be a twice continuously differentiable
function with bounded derivatives, satisfying the Neumann boundary condition
$\partial_{x_k}F(x)\big|_{x_k=0} = 0$ for all $k$. For such $F$, Taylor's
theorem yields, uniformly on compact sets,
\begin{align*}
  F\Big(x \pm \frac{e_k}{\sqrt{n}}\Big)
  &= F(x)
   \pm \frac{1}{\sqrt{n}}\,\partial_{x_k}F(x)
   + \frac{1}{2n}\,\partial^2_{x_kx_k}F(x)
   + O\big(n^{-3/2}\big),\\
  \Delta^r_{n,k}F(x)
  &= \partial_{x_k}F(x) + O\big(n^{-1/2}\big),\\
  \Delta^l_{n,k}F(x)
  &= \partial_{x_k}F(x) + O\big(n^{-1/2}\big),\\
  \Delta^2_{n,k}F(x)
  &= \partial^2_{x_kx_k}F(x) + O\big(n^{-1/2}\big).
\end{align*}
A similar expansion applied to the migration combination
\[
  \Delta^r_{n,k-1}F(x - e_k/\sqrt{n})
  + \Delta^r_{n,k+1}F(x - e_k/\sqrt{n})
  - 2\Delta^l_{n,k}F(x)
\]
shows that
\[
  \Delta^r_{n,k-1}F(x - e_k/\sqrt{n})
  + \Delta^r_{n,k+1}F(x - e_k/\sqrt{n})
  - 2\Delta^l_{n,k}F(x)
  = \big(x_{k+1} + x_{k-1} - 2x_k\big)\,\partial_{x_k}F(x)
    + O\big(n^{-1/2}\big).
\]

\paragraph{Remark 5.2.2}
These Taylor expansions are standard in the analysis of weak convergence
for scaled Markov jump processes \cite[Ch.~7]{ethierkurtz1986}. Very
similar generator expansions appear in the limit order book scaling limits
of Hambly and Ledger [7] and Horst and Paul [6], to which our setting is closely related.

Substituting these expansions into (10), we obtain
\[
  A_nF(x) = AF(x) + R_n(F,x),
\]
where the \emph{candidate limit generator} $A$ is given by
\begin{equation}
  AF(x)
  = \sum_{k=1}^{N-1}
    \bigg[
      \frac{1}{2}\,\sigma_k^2(x_k)\,\frac{\partial^2F}{\partial x_k^2}(x)
      + \Big(
          h_k(x_k)
          + \alpha_b\big(x_{k+1} + x_{k-1} - 2x_k\big)
        \Big)\,\frac{\partial F}{\partial x_k}(x)
    \bigg],
  \label{eq:limit-generator}
\end{equation}
with $h_k := f_k - g_k$, and the remainder satisfies
\[
  \sup_{x \in K}\,\big|R_n(F,x)\big| \;\longrightarrow\; 0
  \qquad \text{for  every compact } K \subset \mathbb{R}_+^{N-1}.
\]

The operator $A$ is understood on the domain
\[
  \mathcal{D}(A)
  := \big\{
      F \in C^2_b(\mathbb{R}_+^{N-1}) :
      \partial_{x_k}F(x)\big|_{x_k=0} = 0
      \text{ for all } k
     \big\},
\]
which encodes reflection at the boundary $x_k = 0$.

\paragraph{Remark 5.2.3}
The interpretation of $A$ as the generator of a diffusion process with
normal reflection on the boundary $\{x_k = 0\}$ is based on the classical
framework of Lions and Sznitman for reflected stochastic differential
equations; see [11].

\subsection{Mesoscopic reflected SDE}

The generator $A$ in (11) is the generator of a
reflected diffusion $X^b = (X^{b,1},\dots,X^{b,N-1})$ solving the SDE system
\begin{equation}
  \begin{aligned}
    dX^{b,i}_t
    &= \Big[
          h_i\big(X^{b,i}_t\big)
          + \alpha_b\big(
              X^{b,i+1}_t + X^{b,i-1}_t - 2X^{b,i}_t
            \big)
       \Big]\,dt \\
    &\quad + \sigma_i\big(X^{b,i}_t\big)\,dW^{b,i}_t
       + d\eta^{b,i}_t,
       \qquad i = 1,\dots,N-1, \\
    X^{b,0}_t &= X^{b,N}_t = 0,
  \end{aligned}
  \label{eq:mesoscopic-SDE}
\end{equation}
where $W^b = (W^{b,1},\dots,W^{b,N-1})$ is a Brownian motion in
$\mathbb{R}^{N-1}$ with covariance structure specified in
Section~\ref{sec:hawkes-covariance}, and $\eta^{b,i}$ are non-decreasing
processes ensuring reflection at zero:
\[
  X^{b,i}_t \ge 0, \quad
  \int_0^\infty X^{b,i}_t\,d\eta^{b,i}_t = 0.
\]

Standard results on martingale problems for reflected diffusions (see,
[11] ) imply that the martingale problem for $(A,\mathcal{D}(A))$
is well posed. Together with the generator convergence $A_n \to A$ on
$\mathcal{D}(A)$ and the general convergence theory for Markov processes in
\cite{ethierkurtz1986}, this yields the following mesoscopic limit:

\begin{theorem}
  As $n \to \infty$, the rescaled queue process $\tilde Z^b_n$ converges in
  distribution in $D([0,\infty);\mathbb{R}_+^{N-1})$ to the reflected diffusion
  $X^b$ solving (12).
\end{theorem}

\section{Diffusion Covariance Induced by Hawkes Structure}
\label{sec:hawkes-covariance}

We now make explicit how the covariance structure of the Brownian motion
$W^b$ in (12) is determined by the underlying Hawkes
kernels.

\subsection{Multivariate Hawkes specification}

We index all microscopic event types on the bid side
by $a = 1,\dots,M$ and collect them in the
$M$-dimensional Hawkes process
\[
  N(t) = \big(N^1(t),\dots,N^M(t)\big)^\top.
\]
Its intensity process $\lambda(t) = (\lambda^1(t),\dots,\lambda^M(t))^\top$
satisfies
\[
  \lambda(t)
  = \mu + \int_0^t \Phi(t-s)\,dN(s),
\]
where $\mu \in \mathbb{R}^M_+$ is the vector of baseline intensities and
$\Phi(t) = \big(\phi_{ij}(t)\big)_{1\le i,j\le M}$ is the non-negative kernel
matrix.

Define the \emph{integrated kernel}
\[
  K := \int_0^\infty \Phi(u)\,du
      = \bigg(\int_0^\infty \phi_{ij}(u)\,du\bigg)_{1\le i,j\le M},
\]
and assume the stability condition $\rho(K) < 1$, where $\rho(K)$ denotes the
spectral radius of $K$. Under stationarity, the mean intensity vector
$\bar\Lambda := \mathbb{E}[\lambda(t)]$ is then given by
\[
  \bar\Lambda = (I - K)^{-1}\mu.
\]

\subsection{Functional central limit theorem for event counts}

The multivariate Hawkes FCLT( see \cite{BacryGaiffasMuzy2015})
 implies that under
the diffusive scaling,
\[
  \frac{1}{\sqrt{n}}\Big(N(nt) - nt\,\bar\Lambda\Big)
  \;\Rightarrow\; \mathcal{N}\big(0,\,t\,\Sigma_N\big),
\]
where the asymptotic covariance matrix $\Sigma_N$ is given by
\begin{equation}
  \Sigma_N
  = (I - K)^{-1}\,\mathrm{diag}(\bar\Lambda)\,(I - K)^{-\top}.
  \label{eq:SigmaN}
\end{equation}
Equivalently, one may write
\[
  \frac{1}{\sqrt{n}}\Big(N(nt) - nt\,\bar\Lambda\Big)
  \;\Rightarrow\;
  (I - K)^{-1}\,\mathrm{diag}(\bar\Lambda)^{1/2}\,W_t,
\]
for a standard $M$-dimensional Brownian motion $W$, in which case
$\Sigma_N = (I - K)^{-1}\mathrm{diag}(\bar\Lambda)(I - K)^{-\top}$ is precisely
the covariance of the Gaussian limit.

\subsection{Covariance of queue volumes}

The (rescaled) bid side queue volumes $\tilde Z^b_n$ are linear functionals of
the event counts $N$. There exists an incidence matrix
$C \in \mathbb{R}^{(N-1)\times M}$ such that
\[
  \tilde Z^b_n(t) = \frac{Z^b_n(nt)}{\sqrt{n}} = C\,\frac{N(nt)}{\sqrt{n}}.
\]
Each column of $C$ encodes the effect of a single event type on the queue
profile (for instance, a unit up-jump at level $i$ contributes $+1$ to row $i$,
a down jump contributes $-1$, and a migration $i \to i+1$ contributes $-1$ to
row $i$ and $+1$ to row $i+1$).

It follows from the FCLT and linearity that
\[
  \frac{Z^b_n(nt) - nt\,C\bar\Lambda}{\sqrt{n}}
  \;\Rightarrow\; \mathcal{N}\big(0,\,t\,\Sigma_X\big),
\]
where the asymptotic covariance matrix of the queue volumes is
\begin{equation}
  \Sigma_X
  = C\,\Sigma_N\,C^\top
  = C\,(I - K)^{-1}\,\mathrm{diag}(\bar\Lambda)\,(I - K)^{-\top}C^\top.
  \label{eq:SigmaX}
\end{equation}
Thus, in the mesoscopic SDE, one can
choose the diffusion matrix $\Gamma(x)$ such that
$\Gamma(x)\Gamma(x)^\top = \Sigma_X$ (for instance, via a Cholesky
factorisation), and the Brownian motion $W^b$ is related to the underlying
Hawkes fluctuations through \eqref{eq:SigmaX}. This makes explicit how
both the endogenous order flow and the self-exciting liquidity migration
mechanisms shape the covariance structure of the mesoscopic queue dynamics.

\section{Conclusions}

In this work we have developed a mesoscopic description of the bid-side
queue dynamics in a limit order book driven by a multivariate Hawkes process.
Starting from a fully microscopic specification in which individual order
arrivals, cancellations and migrations are encoded as components of a Hawkes
counting process, we introduced a diffusive rescaling under which time is
accelerated by a factor $n$ and queue sizes are of order $\sqrt{n}$. In this
regime, and under suitable equilibrium intensity expansions, we derived the
infinitesimal generator of the rescaled queue process and showed that it
converges to the generator of a reflected diffusion on $\mathbb{R}_+^{N-1}$.

More concretely, we obtained a mesoscopic SDE system, in which each queue level evolves as a one-dimensional
diffusion with state dependent drift and volatility, coupled across levels through
a discrete Laplacian term that encodes liquidity migration. Reflection at zero is
implemented via a Skorokhod type term ensuring non-negativity of the queues and a
Neumann boundary condition in the associated generator. This provides a rigorous
link between the underlying self-exciting order flow and an effective
interacting diffusion description of the queue dynamics.

\paragraph{}
In the second part of the analysis we made explicit how the diffusion
coefficients in the mesoscopic model are determined by the Hawkes structure.
Using a functional central limit theorem for multivariate Hawkes processes, we
expressed the asymptotic covariance matrix of the event counts in terms of the
integrated kernel and the stationary intensities. Through a
linear incidence mapping between event types and queue increments, this yields a
closed form expression for the covariance matrix of the mesoscopic queue
fluctuations and, consequently, for the diffusion matrix. In this way, the model retains a clear microscopic
interpretation while admitting a tractable diffusion limit
\subsection{Future Research}

A key next step is to calibrate the microscopic Hawkes parameters and the
resulting mesoscopic coefficients to high frequency limit order book data.
This would involve estimating the kernel matrix and baseline intensities from
event time series, constructing the corresponding branching matrix, and
comparing empirical queue fluctuations with those predicted by the diffusion
limit. One could then assess, for example, how well the Laplacian coupling
captures observed liquidity migration and how the theoretical covariance
structure compares to empirical covariances across levels and maturities.

From a computational perspective, it would be important to develop efficient
numerical schemes for simulating the reflected mesoscopic SDE, taking into account both the non-negativity
constraint and the coupling across levels. This includes the design and
analysis of discretisation methods that preserve reflection and stability,
as well as variance reduction techniques that exploit the Hawkes induced
covariance structure. Such schemes would be useful both for model validation
and for applications in optimal execution, liquidity risk management and
stress testing.
\paragraph{
}
Another direction is to consider a further scaling in which the spacing
between price levels tends to zero, leading to a continuum of levels limit.
In this regime one expects the discrete Laplacian coupling to converge to a
second order spatial derivative, and the mesoscopic SDE system to approach a
stochastic partial differential equation (SPDE) for the queue density as a
function of price and time. This would provide a bridge between microscopic
Hawkes based models and macroscopic SPDE descriptions of limit order books,
and could be used to analyse large scale properties such as liquidity
profiles and volatility clustering.

Finally, the Hawkes specification itself can be enriched in several ways,
for instance by allowing cross-excitation between different levels and
event types, including state dependence in the kernel or baseline terms, or
incorporating regimes and non linear saturation effects. Each of these
extensions would induce a modified diffusion covariance structure and could
lead to qualitatively different mesoscopic behaviour. Understanding the
impact of such modeling choices on the resulting diffusion limit and on
observables such as queue correlation, imbalance dynamics and price impact
remains an interesting topic for future work.

Overall, the mesoscopic reflected diffusion derived here offers a natural
starting point for further theoretical analysis and calibration to market data.


\begin{thebibliography}{99}
\setlength{\itemsep}{1em}

\bibitem{HamblyKalsiNewbury2018}
B.~Hambly, J.~Kalsi, and J.~Newbury,
``Limit order books, diffusion approximations and reflected SPDEs: from microscopic to macroscopic models,''
arXiv:2018.

\bibitem{ParlourSeppi2008}
C.~E. Parlour and D.~J. Seppi,
``Limit order markets: A survey,''
in \textit{Handbook of Financial Intermediation and Banking},
A.~W. Thakor and A.~Boot, Eds.,
Amsterdam, Netherlands: Elsevier, 2008, pp.~63--95.

\bibitem{Hewlett2006}
P.~Hewlett,
``Clustering of order arrivals, price impact and trade path optimisation,''
\textit{Workshop on Financial Modeling with Jump Processes},
Ecole Polytechnique, 2006, pp.~6--8.

\bibitem{ContLarrard2011}
R.~Cont and A.~de~Larrard,
``Price dynamics in a Markovian limit order market,''
arXiv:1104.4596, 2011.

\bibitem{Bacry2015}
E.~Bacry, S.~Gaiffas, and J.~F. Muzy,
``Queue-reactive Hawkes processes for limit order books,''
arXiv:1502.04592, 2015.

\bibitem{HorstPaul2019}
U.~Horst and J.~Paul,
``A diffusion limit for a Hawkes-type limit order book model,''
arXiv:1901.06740, 2019.

\bibitem{HamblyLedger2018}
B.~Hambly and S.~Ledger,
``A continuum limit for order book models,''
\textit{SIAM Journal on Financial Mathematics},
vol.~9, no.~3, pp.~865--911, 2018.

\bibitem{Chen2017}
X.~Chen,
``Modelling order flow dynamics using Hawkes processes,''
Ph.D. dissertation, Florida State University, 2017.

\bibitem{AbergelJedidi2011}
F.~Abergel and A.~Jedidi,
``A mathematical approach to order book modeling,''
arXiv:1104.4596, 2011.
\bibitem{ethierkurtz1986}
S.~N. Ethier and T.~G. Kurtz,
\textit{Markov Processes: Characterization and Convergence},
Wiley, New York, 1986.
\bibitem{LionsSznitman1984}
J.-L. Lions and A.-S. Sznitman,
``Stochastic differential equations with reflecting boundary conditions,''
\textit{Communications on Pure and Applied Mathematics},
vol. 37, no. 4, pp. 511--537, 1984.
\bibitem{BacryGaiffasMuzy2015}
E.~Bacry, S.~Ga\"iffas, and J.-F.~Muzy,
``A generalized method for multivariate Hawkes processes,''
\textit{arXiv preprint} arXiv:1502.04592, 2015.


\end{thebibliography}
\end{document}